\RequirePackage{fix-cm}
\documentclass[twocolumn,epjc3]{svjour3} 
\usepackage{CJK}
\usepackage{widetext}

\usepackage{anyfontsize}
\usepackage{xfrac}
\usepackage{relsize}
\usepackage{tikz}
\usepackage[com pat=1.1.0]{tikz-feynman}
\usetikzlibrary{calc,arrows,decorations.markings,shapes}
\usepackage{pgfplots}
\pgfplotsset{compat=1.3}
\usetikzlibrary{3d,calc}

\newcommand{\crn}{\nonumber \\}

\tikzset{every picture/.style={
    scale=0.5, transform shape,
    }}

\usepackage{units}

\usepackage{xspace}

\usepackage{color}
\usepackage{subcaption}
\usepackage{graphicx}
\usepackage[space]{grffile}
\usepackage{verbatim}
\usepackage{amsmath}
\usepackage{amssymb}
\usepackage{wasysym}
\usepackage{ifsym}%
\usepackage{url}
\usepackage{bbold}
\usepackage{slashed}
\usepackage{epstopdf}
\usepackage{braket}
\usepackage{float}

\newcommand{\boundellipse}[3]
{[black,fill=blue!30] (#1) ellipse (#2 and #3)
}
\newcommand{\boundellipseW}[3]
{[white,fill=white] (#1) ellipse (#2 and #3)
}

\newcounter{countitems}
\newcounter{nextitemizecount}
\newcommand{\setupcountitems}{%
  \stepcounter{nextitemizecount}%
  \setcounter{countitems}{0}%
  \preto\item{\stepcounter{countitems}}%
}
\makeatletter
\newcommand{\computecountitems}{%
  \edef\@currentlabel{\number\c@countitems}%
  \label{countitems@\number\numexpr\value{nextitemizecount}-1\relax}%
}
\newcommand{\nextitemizecount}{%
  \getrefnumber{countitems@\number\c@nextitemizecount}%
}
\newcommand{\previtemizecount}{%
  \getrefnumber{countitems@\number\numexpr\value{nextitemizecount}-1\relax}%
}
\makeatother    
{\end{itemize}%
\unskip\computecountitems\ifnumcomp{\previtemizecount}{>}{3}{\end{multicols}}{}}

\RequirePackage{graphicx}
\RequirePackage{latexsym}
\usepackage{doi}
\RequirePackage[numbers,sort&compress]{natbib}
\usepackage{hyperref}
\journalname{Eur. Phys. J. C}
\begin{document}

\title{Constraints from muon \texorpdfstring{$g-2$}{} on a gauged non-universal \texorpdfstring{$U(1)_{X}$}{U(1)X} model with inverse see-saw neutrinos}

\author{J. S. Alvarado\thanksref{e1,addr1}
        \and
        R. Martinez\thanksref{e2,addr2}
        \and
        Cristian Sierra\thanksref{e3,addr3,addr4}
}

\thankstext{e1}{e-mail: alvarado-galeano@ijclab.in2p3.fr}
\thankstext{e2}{e-mail: remartinezm@unal.edu.co}
\thankstext{e3}{e-mail: cristian.sierra@njnu.edu.cn}

\institute{Université Paris-Saclay, CNRS/IN2P3, IJCLab, 91405 Orsay, France \label{addr1}
           \and
           Departamento de Física, Universidad Nacional de Colombia, Ciudad Universitaria, K. 45 No. 26-85, Bogotá D.C., Colombia \label{addr2}
           \and
           Department of Physics and Institute of Theoretical Physics, Nanjing Normal University, Nanjing, Jiangsu 210023, China \label{addr3}
           \and
           Tsung-Dao Lee Institute \& School of Physics and Astronomy, Shanghai Jiao Tong University, Shanghai 200240, China \label{addr4}
}

\date{Received: date / Accepted: date}
\maketitle

\begin{abstract}

We study the effects on a non-universal $U(1)_{X}$ extension of the Standard Model given the alternative value obtained by the Budapest-Marseille-Wuppertal (BMW) group for the anomalous magnetic moment of the muon $g-2$. The model explains the fermion mass hierarchy through the non-universality of the extra gauge symmetry and by an additional $\mathbb{Z}_{2}$ discrete symmetry, where the heaviest fermions acquire their masses from two different scales determined by two Higgs doublets and one singlet, whereas the lightest fermions obtain their masses from radiative corrections. From cancellation of chiral anomalies, the model also includes heavy extra fermions, both charged and neutral. The latter are right-handed neutrinos that acquire masses via an inverse see-saw mechanism, reproducing the observed squared mass differences for the active neutrinos. Using the latest lattice calculation of the leading hadronic vacuum polarization (HVP) contribution to the muon $g-2$, we compute the dominant one-loop diagrams mediated by the $W$ and charged Higgs bosons, both with a heavy Majorana neutrino in the loop, setting bounds for masses of the new particles. We also provide predictions for observables that can probe our model in the future such as charged lepton flavor violating searches at Belle II like $\tau\to \mu\gamma$, $\tau\to e\gamma$ and at MEG II for $\mu\to e\gamma$.

\end{abstract}

\maketitle	

\section{Introduction}
From a phenomenological point of view the Standard Model (SM), although greatly successful, is still an incomplete model from both theoretical and experimental reasons. One of the theoretical reasons is that it cannot accommodate the observed mass hierarchy of its fermionic particle content, which is often called the flavor puzzle. On the experimental side,  it has not been able to explain several persistent observations from both flavor and lepton flavor universality violation searches in colliders. Namely, the observed deviations from the SM in the measurements of flavor changing charged currents (FCCCs) of around $3.2\sigma$ from the SM~\cite{HFLAV:2022esi} for the $R(D^{(*)})$ ratios and on the other hand, measurements of flavor changing neutral currents (FCNCs) close to a combined $6\sigma$ deviation in $b\to s\ell\ell$ observables~\cite{Descotes-Genon:2015uva,Capdevila:2017bsm,Hurth:2020rzx,Bhom:2020lmk,Alguero:2021anc,Hurth:2021nsi}, hint together at the existence of models including new physics (NP) contributions, serving as a clear motivation for SM extensions.

One of such common models extending the SM is the Two Higgs Doublet Model (2HDM), which can fit the charged current anomalies at the $1\sigma$ level \cite{Iguro:2018qzf,Athron:2021auq,Crivellin:2023sig}.
Similar analyses for the neutral current anomalies have also been addressed by previous works \cite{Iguro:2018qzf,Crivellin:2019dun,Crivellin:2023sig}, finding solutions at the $2\sigma$ level and down to the $1\sigma$ level once TeV-scale right-handed neutrinos are included \cite{Crivellin:2019dun}, addressing on top of those the anomalous magnetic moment of the muon $g-2$ simultaneously. \cite{Athron:2024rir,Athron:2024mcn}.

The flavor conserving anomalous magnetic moment of the muon, $a_\mu\equiv(g-2)/2$ is an important observable for NP \cite{Stockinger:2006zn,Athron:2021iuf} which is often considered alongside flavor violating observables in phenomenological studies. On the experimental side the Fermilab measurement of the anomalous magnetic moment of the muon \cite{Muong-2:2023cdq,PhysRevLett.126.141801} combined with results from the Brookhaven National Laboratory give a measured value of $a_{\mu}^{\textrm{Exp}} = 11659205.9\pm2.2\times10^{-10}$. This is larger than the SM value computed by the $g-2$ Theory Initiative's White Paper 
(WP) \cite{Aoyama:2020ynm}, $a_{\mu}^{\textrm{WP}} = 11659181.0\pm4.3\times10^{-10}$, which is based on work from \cite{davier:2017zfy,keshavarzi:2018mgv,colangelo:2018mtw,hoferichter:2019gzf,davier:2019can,keshavarzi:2019abf,kurz:2014wya,melnikov:2003xd,masjuan:2017tvw,Colangelo:2017fiz,hoferichter:2018kwz,gerardin:2019vio,bijnens:2019ghy,colangelo:2019uex,colangelo:2014qya,Blum:2019ugy,aoyama:2012wk,Aoyama:2019ryr,czarnecki:2002nt,gnendiger:2013pva}, giving $\Delta a_{\mu}^\textrm{WP} = (24.9\pm4.8)\times10^{-10}$.

However, this apparent discrepancy of $5.1\sigma$ depends crucially upon the data driven estimates (based on dispersion relations and $e^+e^- \to \pi^+ \pi^-$ data)  of the leading hadronic vacuum polarization (HVP) contributions to muon $g-2$ that were used in the WP.  This WP prediction does not include any lattice calculations of the HVP contribution, but around the same time as the WP the Budapest-Marseille-Wuppertal (BMW) collaboration \cite{Borsanyi:2020mff,Boccaletti:2024guq}, published the first lattice determination of the HVP contribution in the SM with an uncertainty estimate very competitive with the data driven estimates. This BMW prediction $a_{\mu}^{\textrm{BMW}} = (11691954\pm58)\times10^{-11}$ is not in agreement with those data driven estimates used in the WP, and is significantly larger such that using this instead the WP gives a remarkably small deviation ($0.9 \sigma$ from the SM, i.e., in agreement with it) from experiment, $\Delta a_{\mu}^\textrm{BMW} = (4.0\pm4.4)\times10^{-10}$.

Recent results by other lattice groups have found agreement with this \cite{Ce:2022kxy,ExtendedTwistedMass:2022jpw,RBC:2023pvn,FermilabLatticeHPQCD:2023jof} in an intermediate Euclidean time window  \cite{RBC:2018dos}, where lattice systematics are smaller.  This shows a consistent picture of a tension between lattice and the data driven result for this window observable \cite{Colangelo:2022vok}.  The situation is further complicated by the the data from CMD\cite{CMD-3:2023alj}, which gives systematically lower cross sections than BaBar \cite{BaBar:2012bdw} and KLOE \cite{KLOE-2:2017fda} for $e^+e^-\rightarrow \pi^+\pi^-$. The implications of these data are being investigated in detail \cite{Davier:2023fpl}, while the exact origin of the discrepancy between the lattice and the data-driven estimates is still being investigated \cite{Davier:2023cyp}, but if the CMD-3 data are correct it could lead to resolution of the discrepancy \cite{Benton:2023dci}. Finally, $\tau$-data-driven results also seem to be consistent with the lattice results \cite{Masjuan:2023qsp,Davier:2023fpl}. 

In this paper, we will use the model presented in~\cite{orig}, which is a gauged non-universal $U(1)_X$ extension of the 2HDM. The model explains the mass hierarchy of charged fermions and neutrino oscillations thanks to the non-universality of the new quantum numbers and a $\mathbb{Z}_{2}$ symmetry. The lightest fermions, (i.e., the down and strange quarks, and the electron) can obtain their masses radiatively \cite{Garnica:2019hvn} or by the Froggatt-Nielsen mechanism \cite{Alvarado:2021nxy,Froggatt:1978nt}, while neutrino masses are explained by an inverse see-saw mechanism \cite{orig}. The model can also fit the $B$ meson flavor anomalies at the $2\sigma$ level as we showed in our previous work \cite{Alvarado:2021nxy}. New interactions arise from the mixing between the SM particles with heavy exotic fermions required by cancellation of chiral anomalies \cite{Ellis:2017nrp,Allanach:2018vjg}. In particular we will show what the constraints on the masses of such extra heavy particles are by using the most stringent prediction on $\Delta a_{\mu}$ from the BMW lattice group updating the results of our previous work \cite{Alvarado:2021nxy} and obtaining predictions for charged flavor violating (cLFV) decays at Belle II and MEG II.

\section{The \texorpdfstring{$U(1)_{X}$}{U(1)X} extension}\label{modelgen}

\begin{table}
\centering %
\begin{tabular}{|ccc|}
\hline 
Scalars  & $X^{\pm}$  & $Y$ \tabularnewline
\hline 
\hline 
$\phi_{1}=\begin{pmatrix}\phi_{1}^{+}\\
\frac{h_{1}+v_{1}+i\eta_{1}}{\sqrt{2}}
\end{pmatrix}$  & $\sfrac{+2}{3}^{+}$  & $+1$ \tabularnewline
$\phi_{2}=\begin{pmatrix}\phi_{2}^{+}\\
\frac{h_{2}+v_{2}+i\eta_{2}}{\sqrt{2}}
\end{pmatrix}$ & $\sfrac{+1}{3}^{-}$ & $+1$\tabularnewline
$\chi=\frac{\xi_{\chi}+v_{\chi}+i\zeta_{\chi}}{\sqrt{2}}$  & $\sfrac{-1}{3}^{+}$  & $0$ \tabularnewline
\hline 
\end{tabular}\caption{Scalar particle content of the model, $X$ charge, $\mathbb{Z}_{2}$
parity ($\pm$) and hypercharge $Y$.}
\label{scalarlist} 
\end{table}

The scalar sector of the model consists of two doublets $\phi_{1,2}$ with Vacuum Expectation Values (VEV) $\upsilon_{1,2}$  related as $\upsilon=\sqrt{\upsilon_1^2+\upsilon_2^2}=246\,\mathrm{GeV}$. Additionally, the model includes one scalar singlet $\chi$ associated with the spontaneous symmetry breaking (SSB) of the $U(1)_{X}$ symmetry via its respective VEV, $v_{\chi}$, at the TeV scale. We summarize the scalar particle content and its quantum numbers in Table \ref{scalarlist}. 

The most general renormalizable scalar potential of the model consistent with the symmetries in Table \ref{scalarlist} is given by \cite{orig},

\begin{align}
V &= \mu_{1}^{2}\phi_{1}^{\dagger}\phi_{1} + \mu_{2}^{2}\phi_{2}^{\dagger}\phi_{2} + \mu_{\chi}^{2}\chi^{*}\chi + \frac{f}{\sqrt{2}}\left(\phi_{1}^{\dagger}\phi_{2}\chi ^{*} + \mathrm{\text{h.c.}} \right)  \nonumber\\ &
 + \lambda_{1}\left(\phi_{1}^{\dagger}\phi_{1}\right)^{2} + \lambda_{2}\left(\phi_{2}^{\dagger}\phi_{2}\right)^{2} 
 + \lambda_{3}\left(\chi^{*}\chi \right)^{2}  \nonumber\\	&
 + \lambda_{5}\left(\phi_{1}^{\dagger}\phi_{1}\right) \left(\phi_{2}^{\dagger}\phi_{2}\right)
 + \lambda'_{5}\left(\phi_{1}^{\dagger}\phi_{2}\right)\left(\phi_{2}^{\dagger}\phi_{1}\right)	\nonumber\\ 
 & + \lambda_{6} \left(\phi_{1}^{\dagger}\phi_{1}\right) \left(\chi^{*}\chi \right)   + \lambda_{7}\left(\phi_{2}^{\dagger}\phi_{2}\right) \left(\chi^{*}\chi \right).
\end{align}

After SSB of the $U(1)_{X}$ and electroweak symmetries, this potential generates mass matrices for the charged, CP-even and CP-odd scalars (see Table \ref{table_masses}).  The lightest CP-even state, whose mass is proportional to the electroweak VEV, is identified with the Higgs boson. The additional scalars of the model correspond to heavy mass states depending on $v_{\chi}$.

\begin{table}[h]
    \centering
    \begin{tabular}{|cc|} \hline
        Physical Scalars & Mass Eigenvalue  \\ \hline \hline
        $H^{\pm}$ (Charged scalar) & $-\frac{1}{4}\frac{f v_{\chi}}{s_{\beta}c_{\beta}} -\frac{1}{4}\lambda_{5}' v^2$  \\
        $A^{0}$ (Pseudo-scalar) & $-\frac{1}{4}\frac{f v_{\chi}}{s_{\beta}c_{\beta}c_{\gamma}^{2}} \approx -\frac{1}{4}\frac{f v_{\chi}}{s_{\beta}c_{\beta}}$  \\
        $h$ (SM Higgs bosons) & $\left( \tilde{\lambda}_{1}c_{\beta}^{4}+2\tilde{\lambda}_{5}c_{\beta}^{2}s_{\beta}^{2}+\tilde{\lambda}_{2}s_{\beta}^{4} \right){v^{2}}$ 	\\
        $H$ (Heavy scalar) & $-\frac{f v_{\chi}}{4s_{\beta}c_{\beta}}$ 	\\
        $H_{\chi}$ (Heavy scalar) & $\lambda_{3} {v_{\chi}^{2}}$  \\ \hline \hline
        Physical Leptons & Mass Eigenvalue  \\ \hline \hline
        $e$ & $0$ \\
        $\mu$ & $\frac{1}{2}(\eta^{2} + h^{2})v_{2}^{2}$ \\
        $\tau$ & $\frac{1}{2} H^{2}v_{2}^{2}$ \\
        $\nu^{e,\mu,\tau}$ (SM neutrino) & $\propto \frac{\mu_{N} v_{2}^{2}}{{h_{N\chi e}}^{2}v_{\chi}^{2}}$ \\
        $E$ (Exotic lepton) & $\frac{1}{2}g_{\chi E}^{2}v_{\chi}^{2}$ \\    
    $\mathcal{N}^{i}$ (Exotic neutrino) & $\frac{1}{2}\,(\,\mu_{N}\mp\sqrt{\mu_{N}^{2}+2h_{N_{\chi i}}^{2}v_{\chi}^{ 2}}\,)$ \\ \hline
    \end{tabular}
    \caption{Mass eigenvalues of the relevant particle content. We define $s_{\beta}=\sin\beta$, $c_{\beta}=\cos\beta$, $t_{\beta}=s_{\beta}/c_{\beta}=v_{1}/v_{2}$ with $v_{1}>v_{2}$ and $t_{\gamma}=\tan\gamma=v s_{\beta}c_{\beta}/v_{\chi} \ll 1$. The constants with a tilde are defined in \cite{Alvarado:2021nxy}. For the heavy neutrino masses $i=1,...,6$, the first sign in $\mp$ corresponds to  $i=1,2,3$.}
    \label{table_masses}
\end{table}

In the leptonic sector, the SM particle content is extended with two exotic charged lepton singlets $E$ and $\mathcal{E}$, three right-handed neutrinos $\nu_{R}$ and three Majorana neutrinos $\mathcal{N}_{R}$. The leptonic particle content is summarized in Table \ref{fermionlist}. The mass eigenvalues for the leptons are summarized in Table \ref{table_masses}, where all parameters different from the VEVs are the $\mathcal{O}(1)$ interaction couplings entering the charged leptons Lagrangian,

\begin{align}
-\mathcal{L}_{\ell} & =\eta\overline{\ell_{L}^{e}}\phi_{2}e_{R}^{\mu}+h\overline{\ell_{L}^{\mu}}\phi_{2}e_{R}^{\mu}+\zeta\overline{\ell_{L}^{\tau}}\phi_{2}e_{R}^{e}\nonumber \\
 & +H\overline{\ell_{L}^{\tau}}\phi_{2}e_{R}^{\tau}+q_{11}\overline{\ell_{L}^{e}}\phi_{1}E_{R}+q_{21}\overline{\ell_{L}^{\mu}}\phi_{1}{E}_{R}\nonumber \\
 & +g_{\chi E}\overline{E_{L}}\chi E_{R}+g_{\chi\mathcal{E}}\overline{\mathcal{E}_{L}}\chi^{*}\mathcal{E}_{R}+\mathrm{\text{h.c.}}.
\end{align}

Among the exotic charged leptons, only $E$ couples to the SM particles through the interaction with electrons and muons, resulting in the following fermionic mass matrix,
\begin{equation}
\mathbb{M}_{E}^{0} = 
\frac{1}{\sqrt{2}}\left(\begin{array}{ c c c |c c}
    0 & \eta\,v_{2} & 0 &  q_{11}\,v_{1}   \\
    0 & h\,v_{2}  & 0 &  q_{12}\,v_{1}  \\
    \zeta\,v_{2}  & 0 & H\,v_{2} & 0 \\ \hline
    0 & 0 & 0 & g_{\chi E}\,v_{\chi}
    \end{array} \right).
\end{equation}
Mass eigenvalues are obtained by squaring the mass matrix, where the $\tau$ lepton decouples and the light leptons can be obtained by noticing that the $E$ mass scale is much larger than the SM lepton masses.

\begin{table}[h]
\centering
\begin{tabular}{|ccc|ccc|}
\hline
SM 	&	$X$ & $\mathbb{Z}_{2}$ & Non-SM  & $X$ & $\mathbb{Z}_{2}$	\\ \hline \hline
$\ell^{e}_{L}=\left(\begin{array}{c}\nu^{e} \\ e^{e} \end{array}\right)_{L}$
	&	$0$	&$+$ &	$\nu_{R}^{e,\mu,\tau}$ & $1/3$ &  $-$\\
$\ell^{\mu}_{L}=\left(\begin{array}{c}\nu^{\mu} \\ e^{\mu} \end{array}\right)_{L}$
	&	$0$	&$+$ & $\mathcal{N}_{R}^{e,\mu,\tau}$ & $0$ & $-$	\\
$\ell^{\tau}_{L}=\left(\begin{array}{c}\nu^{\tau} \\ e^{\tau} \end{array}\right)_{L}$
	&	$-1$	&$+$ & $E_{L},\mathcal{E}_{R}$	&	$-1$	&$+$\\ 
$e_{R}^{e,\tau}$ & $-4/3$ & $-$ & $\mathcal{E}_{L},E_{R}$	&	$-2/3$	&$+$ \\
$e_{R}^{\mu}$ & $-1/3$ & $-$ & & & \\ \hline
\end{tabular}
\caption{Lepton particle content of the model, $X$ charge, $\mathbb{Z}_{2}$ parity and hypercharge. These particular solutions are given by integer values of an elementary $X_{0}=1/3$ charge. Exotic particles are also introduced in the quark sector but are not discussed in this work. See \cite{Alvarado:2021nxy} for more details.}
\label{fermionlist}
\end{table}

Regarding the neutral leptons, the respective Yukawa Lagrangian is given by
\begin{align}
    \mathcal{L}_{NL}&= h_{2p}^{\nu w} \bar{\ell}_{L}^{p}\tilde{\phi}_ {2}\nu_{R}^{w} + h_{\chi \omega}^{\nu j}\bar{\nu}_{R}^{w\; c}\chi^{*}N_{R}^{j} \nonumber \\
    &+ \frac{1}{2}\mu_{N}\delta^{ij}\bar{N}_{R}^{i\; c}N_{R}^{j}\label{lagN},
\end{align}

and $p=e,\mu$ label the lepton doublets, $w=e,\mu,\tau$ for right-handed neutrinos and $i,j=e,\mu,\tau$ for the Majorana neutrinos. The neutrino mass matrix is a $9\times 9$ matrix written in the basis $\left(\begin{matrix}{\nu^{e,\mu,\tau}_{L}},\,\left(\nu^{e,\mu,\tau}_{R}\right)^{C},\,\left(N^{e,\mu,\tau}_{R}\right)^{C}\end{matrix}\right)$,

\begin{align}
\mathcal{M}_{\nu} &=\begin{pmatrix}
    0 & m_{D}^{T} & 0 \\
    m_{D} & 0 & M_{D}^{T} \\
    0 & M_{D} & M_{M}
    \end{pmatrix},     
\end{align}

where the block matrices are defined as,
\begin{align}
 m_{D}^{T}&=\frac{v_{2}}{\sqrt{2}}\begin{pmatrix}
    h_{2e}^{\nu e} & h_{2e}^{\nu \mu} & h_{2e}^{\nu \tau} \\
    h_{2\mu}^{\nu e} & h_{2\mu}^{\nu \mu} & h_{2\mu}^{\nu \tau} \\
    0 & 0 & 0
    \end{pmatrix},
    \label{eq:Dirac_matrix}
\end{align}

and $(M_{D})^{ij}=(1/\sqrt{2}\,)v_{\chi}\,h_{\chi i}^{\nu j}$, $(M_{M})_{ij}=\mu_{N} \mathbb{I}_{3\times 3}$. Neutrino masses are generated via an inverse see-saw mechanism (see \cite{Alvarado:2021nxy,Hong:2023rhg}) by assuming the hierarchy $M_{M}\ll m_{D} \ll M_{D}$. Following \cite{Hong:2023rhg}, the neutrino mixing matrix $U^{\nu}$ can be written as
\begin{align}	
	U^\nu=& \begin{pmatrix}(I_3 - \dfrac{1}{2} RR^\dagger)U_{\mathrm{PMNS}} & RV\\ -R^\dagger U_{\mathrm{PMNS}} & (I_K - \dfrac{1}{2}R^\dagger  R)V \end{pmatrix} \nonumber \\
    &+\mathcal{O}(R^3),
\end{align}

\noindent where $U_{\mathrm{PMNS}}$ is the Pontecorvo-Maki-Nakagawa-Sakata neutrino matrix. The $V$ and $R$ block matrices given by \cite{Thao:2017qtn}

\begin{align}
V&\simeq  \frac{1}{\sqrt{2}}  \begin{pmatrix}
	-iI_3& I_3 \\
	iI_3& I_3
\end{pmatrix},\\
R &\simeq \left(\mathbb{0}_{3\times3},\;  U_{\mathrm{PMNS}}\,\left(\frac{\hat{m}_\nu}{\mu_N}\right)^{1/2} \right).
\end{align}

Given that the model predicts one active neutrino to be massless, we define $\hat{m}_{\nu}=\textrm{diag}(0,\sqrt{\Delta m_{21}^2},\sqrt{\Delta m_{31}^2})$ for the diagonal matrix of the three active neutrinos in the normal ordering scheme, where $\Delta m_{i1}^2$ are the current observed squared mass differences $\Delta m_{21}^2=7.49\times10^{-5}\,\mathrm{eV}^{2}$ and $\Delta m_{31}^2=2.534\times10^{-3}\,\mathrm{eV}^{2}$\cite{Esteban:2024eli}. The masses of the eigenstates of exotic neutrinos $\mathcal{N}^{k}$, $k=1,...,6.$ are given at the $\mu_{N}$ scale. This scale is present in the model as a free parameter with the only condition that the ratio $\mu_{N}v_{2}^{2}/v_{\chi}^{2}$ sets the active neutrinos mass scale. 

Therefore, the mass-eigenvalue spectrum is governed by the different energy scales on the model, leading to six nearly degenerated Majorana neutrinos and non-SM scalars at TeV scale. At the same time, the SM mass hierarchy is explained by the presence of the two electroweak VEVs $v_{1}$ and $v_{2}$ \cite{orig}. 

The above-mentioned particle content and charges arise as a solution of the triangle anomalies, 

{\small{}
\begin{align}
\left[\mathrm{\mathrm{SU}(3)}_{C}\right]^{2}\mathrm{\mathrm{U}(1)}_{X}:\,A_{C}&=  \sum_{Q}\left[X_{Q_{L}}-X_{Q_{R}}\right],\\
\left[\mathrm{\mathrm{SU}(2)}_{L}\right]^{2}\mathrm{\mathrm{U}(1)}_{X}:\,A_{L}&= \sum_{\ell}\left[X_{\ell_{L}}+3X_{Q_{L}}\right],\nonumber \\
\left[\mathrm{\mathrm{U}(1)}_{Y}\right]^{2}\mathrm{\mathrm{U}(1)}_{X}:\,A_{Y^{2}}&= \sum_{\ell,Q}\left[Y_{\ell_{L}}^{2}X_{\ell_{L}}+3Y_{Q_{L}}^{2}X_{Q_{L}}\right]\nonumber \\
 & -\sum_{\ell,Q}\left[Y_{\ell_{R}}^{2}X_{L_{R}}+3Y_{Q_{R}}^{2}X_{Q_{R}}\right],\nonumber \\
\mathrm{\mathrm{U}(1)}_{Y}\left[\mathrm{\mathrm{U}(1)}_{X}\right]^{2}:\,A_{Y}&=  \sum_{\ell,Q}\left[Y_{\ell_{L}}X_{\ell_{L}}^{2}+3Y_{Q_{L}}X_{Q_{L}}^{2}\right]\nonumber \\
 & -\sum_{\ell,Q}\left[Y_{\ell_{R}}X_{\ell_{R}}^{2}+3Y_{Q_{R}}X_{Q_{R}}^{2}\right],\nonumber \\
\left[\mathrm{\mathrm{U}(1)}_{X}\right]^{3}:\,A_{X}= & \sum_{\ell,Q}\left[X_{\ell_{L}}^{3}+3X_{Q_{L}}^{3}-X_{\ell_{R}}^{3}-3X_{Q_{R}}^{3}\right],\nonumber \\
\left[\mathrm{Grav}\right]^{2}\mathrm{\mathrm{U}(1)}_{X}:\,A_{\mathrm{G}}= & \sum_{\ell,Q}\left[X_{\ell_{L}}+3X_{Q_{L}}-X_{\ell_{R}}-3X_{Q_{R}}\right], \nonumber
\end{align}
\label{eq:Chiral_anomalies}
}{\small\par}

\noindent where the second equation runs over $SU(2)$ doublets, $Q$ and $\ell$ runs over all quarks and leptons, respectively, and $Y$ is the corresponding weak hypercharge. The electric charge follows the Gell-Mann-Nishijima relationship $Q=\sigma_{3}/2+Y/2$. As a result of the non-universal $X$ charges on the particle content, along with the $\mathbb{Z}_{2}$ symmetry, the fermion mass matrices are provided with a texture that can provide a natural realization of the fermion mass hierarchy.

\section{Observables}\label{g-2anom}
\subsection{$\ell\to\ell'\gamma$ decays}

The branching ratios for the cLFV decays are given by \cite{Lavoura:2003xp,Hue:2017lak,Crivellin:2018qmi}
\begin{align}
	\label{eq_brebaga}
	\mathrm{Br}(\ell_b\to \ell_a\gamma)= \frac{48\pi^2}{G_F^2 m_b^2}\left( \left| c_{(ab)R}\right|^2 + \left| c_{(ba)R}\right|^2\right)\nonumber\\ \times\mathrm{BR}(\ell_b\to \ell_a \overline{\nu_a}\nu_b),\qquad\qquad
\end{align}

where following Ref.\cite{Hong:2023rhg}, the $c_{(ab)R}$ coefficients are expressed as 
\begin{align}
	\label{eq_cabR}	
	c_{(ab)R}&=  c^{H^\pm}_{(ab)R}+ c^{W^\pm}_{(ab)R},\quad c_{(ba)R}= c_{(ab)R}[a \to b,\; b\to a], \nonumber\\
    c^{H^\pm}_{(ab)R}& = \frac{g^2e\;}{32 \pi^2 m^2_W\,m^2_{H^{\pm}} } \,\sum_{i=1}^{9}\left[ \lambda^{L}_{ai } \lambda^{R}_{bi }m_{n_i} f_{\Phi}(x_{i})\right. \nonumber\\
    & \quad +\left.\left( m_{b} \lambda^{L}_{ai } \lambda^{L}_{bi } + m_{a} \lambda^{R}_{ai } \lambda^{R}_{bi }\right)  \tilde{f}_{\Phi}(x_{i}) \right],
\crn 	c_{(ab)R}^{W^\pm}& \simeq\frac{e G_F\,m_{b}}{4\sqrt{2} \pi^2}   \left[  -\frac{5\delta_{ab}}{12} +  \left(U_{\mathrm{PMNS}}\frac{\hat{m}_{\nu}}{\mu_N}U_{\mathrm{PMNS}}^\dagger \right)_{ab}\right.\nonumber \\
& \quad\left.\times \left(\tilde{f}_V \left(x_W\right)+ \frac{5}{12}\right) \right]
\end{align}
with $x_{i}\equiv m^2_{n_i}/m^2_{H^{\pm}}$, $x_W=m_{\mathcal{N}}^2/m_W^2$ while $m_{n_i}\simeq m_{\mathcal{N}}$ for $i= 4,...,9$, i.e., as mentioned above, we consider the heavy Majorana neutrino masses to be degenerated. The different loop functions are given by \cite{Crivellin:2018qmi}
\begin{align}
	\label{eq_fphiX}	
	f_\Phi (x)&=\frac{x^2-1 -2x\ln x}{4(x-1)^3},\\
	\tilde{f}_\Phi(x)&= \frac{2x^3 +3x^2 -6x +1 -6x^2 \ln x}{24(x-1)^4},
\crn 	\tilde{f}_V(x) &= \frac{-4x^4 +49x^3 -78 x^2 +43x -10 -18x^3\ln x}{24(x-1)^4}.
\end{align}
The $\lambda^{L(R)}_{ai}$ couplings that appear above are given by

\begin{align}
	\lambda^{L}_{ai}=& -t_{\beta} \sum_{b=1}^3U^{\nu}_{(b+3)i} (m_D)_{ba},
	%
	\quad \lambda^{R}_{ai}=  -U^{\nu*}_{ai}m_a t_{\beta}^{-1},
    \label{lambda_couplings}
\end{align}
where the Dirac matrix $m_D$ is given by the texture in Eq.(\ref{eq:Dirac_matrix}).

\subsection{Muon $g-2$}
\begin{figure}[ht]
    \LARGE
     \centering
     \begin{subfigure}[b]{0.2\textwidth}
         \centering
             \begin{tikzpicture}
  \begin{feynman}
    \vertex (a) ;
    \vertex [below right = of a] (b);
    \vertex [below left =of a] (c) ;
    \vertex [right = of a] (x){\( f\)};
    \vertex [left =of a] (y){\( f\)} ;
    \vertex [below = of a ](z){\( Z_{2} \)};
    \vertex [above = of a ] (d){\( \gamma \) };
    \vertex [below right = of b] (e){\(\mu \) } ;
    \vertex [below left = of c] (f){\( \mu \) } ;
    \diagram* {
      (a) -- [anti fermion ] (b) -- [boson] (c) -- [anti fermion] (a),
      (a) -- [boson] (d),
      (b) -- [ anti fermion] (e),
      (c) -- [ fermion] (f),
    };
  \end{feynman}
\end{tikzpicture}
         \caption{}
         \label{g-2a}
     \end{subfigure}
     \hfill
      \begin{subfigure}[b]{0.2\textwidth}
    \centering
    \begin{tikzpicture}
  \begin{feynman}
    \vertex (a) ;
    \vertex [below right = of a] (b);
    \vertex [below left =of a] (c) ;
    \vertex [right = of a] (x){\( f\)};
    \vertex [left =of a] (y){\( f \)} ;
    \vertex [below = of a ](z){\( H,A_0 \)};
    \vertex [above = of a ] (d){\( \gamma \) };
    \vertex [below right = of b] (e){\( \mu \) } ;
    \vertex [below left = of c] (f){\( \mu \) } ;
    \diagram* {
      (a) -- [anti fermion ] (b) -- [scalar] (c) -- [anti fermion] (a),
      (a) -- [boson] (d),
      (b) -- [anti fermion] (e),
      (c) -- [ fermion] (f),
    };
  \end{feynman}
\end{tikzpicture}
         \caption{}
         \label{g-2b}
     \end{subfigure}
     \hfill
          \begin{subfigure}[b]{0.2\textwidth}
         \centering
             \begin{tikzpicture}
  \begin{feynman}
    \vertex (a) ;
    \vertex [below right = of a] (b);
    \vertex [below left =of a] (c) ;
    \vertex [right = of a] (x){\( W^+ \)};
    \vertex [left =of a] (y){\( W^- \)} ;
    \vertex [below = of a ](z){\( \mathcal{N} \)};
    \vertex [above = of a ] (d){\( \gamma \) };
    \vertex [below right = of b] (e){\( \mu \) } ;
    \vertex [below left = of c] (f){\( \mu \) } ;
    \diagram* {
      (a) -- [boson ] (b) -- [fermion] (c) -- [boson] (a),
      (a) -- [boson] (d),
      (b) -- [ anti fermion] (e),
      (c) -- [fermion] (f),
    };
  \end{feynman}
\end{tikzpicture}
         \caption{}
         \label{g-2c}
     \end{subfigure}
     \hfill
     \begin{subfigure}[b]{0.2\textwidth}
         \centering
             \begin{tikzpicture}
  \begin{feynman}
    \vertex (a) ;
    \vertex [below right = of a] (b);
    \vertex [below left =of a] (c) ;
    \vertex [right = of a] (x){\( H^+ \)};
    \vertex [left =of a] (y){\( H^- \)} ;
    \vertex [below = of a ](z){\( \mathcal{N} \)};
    \vertex [above = of a ] (d){\( \gamma \) };
    \vertex [below right = of b] (e){\( \mu \) } ;
    \vertex [below left = of c] (f){\( \mu \) } ;
    \diagram* {
      (a) -- [scalar ] (b) -- [fermion] (c) -- [scalar] (a),
      (a) -- [boson] (d),
      (b) -- [ anti fermion] (e),
      (c) -- [fermion] (f),
    };
  \end{feynman}
\end{tikzpicture}
         \caption{}
         \label{g-2d}
     \end{subfigure}
    \caption{One-loop contributions to muon $g-2$ from (a) $Z_2$ neutral gauge boson, (b)  (pseudo)scalars with $f$ being any lepton, (c) $W^{+}$ gauge boson and (d) charged scalars with with exotic neutrinos $\mathcal{N}$.}
    \label{g-2feynman}
\end{figure}
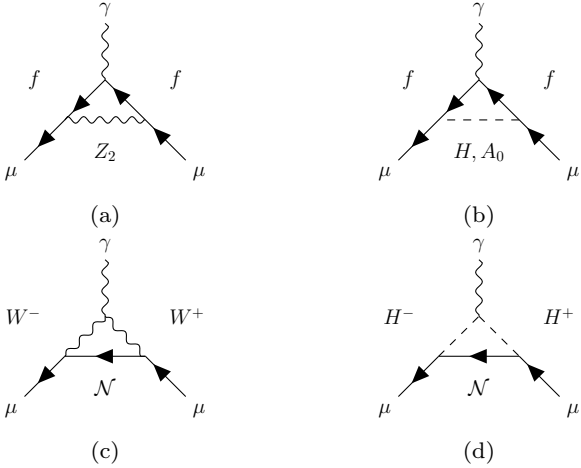

When considering heavy scalars and exotic leptons in the theory, new contributions to the muon $g-2$ value arise, mostly from flavor-changing interactions involving both heavy neutral and charged bosons with those extra leptons as we show in Fig.\ref{g-2feynman}. 

Although new flavor-conserving interactions are also present, their contribution is suppressed by the rotation matrix. Therefore, the diagonal elements of the rotation matrix determine the main contributions to the muon $g-2$, and as shown in \cite{Alvarado:2021nxy}, the dominant one-loop contributions come from the diagrams in Fig.\ref{g-2feynman} (c) and (d), i.e., from the $W^{\pm}$ and $H^{\pm}$ bosons, leading to the following interaction Lagrangians in the mass basis,

\begin{align}
\mathcal{L}_{\mu H^{\pm}\mathcal{N}}= & \frac{-g}{\sqrt{2}\,m_{W}}\bar{\mu}\left(\lambda_{ai}^{L}P_{L}+\lambda_{ai}^{R}P_{R}\right)\mathcal{N}_{i}H^{-}+\text{h.c.},\label{Hpm_Lagrangian}\\
\mathcal{L}_{\mu W\mathcal{N}}= & \frac{-g}{\sqrt{2}}\,\bar{\mu}\,U_{\mu i}^{\nu}P_{L}\,\mathcal{N}_{i}\slashed{W}_{\mu}^{-}+\text{h.c.},\nonumber \\
= & \frac{-g}{\sqrt{2}}\,\bar{\mu}\left(\sqrt{\hat{m}_{\nu}\mu_{N}^{-1}}U_{\text{PMNS}}^{\dagger}\right)_{\mu i}P_{L}\,\mathcal{N}_{i}\slashed{W}_{\mu}^{-}+\text{h.c.},\label{W_Lagrangian}
\end{align}   

where $g$ is the electroweak coupling, the $\lambda_{ai}^{L,R}$ couplings are given by Eq.(\ref{lambda_couplings}) and we have expressed the $\mu W\mathcal{N}$ interaction in terms of the $U_{PMNS}$ matrix and the inverse see-saw scale $\mu_N$. It is worth noting that unlike the $W$ boson contribution, the charged Higgs one will depend explicitly on the Yukawa couplings appearing in the Dirac matrix $m_D$, namely $h_{2j}^{\nu\,i}$, via the $\lambda_{ai}^{L,R}$ couplings.

In this way, the dominant NP contributions to the muon $g-2$ are expressed as \cite{Hong:2023rhg},
\begin{align}
	\label{eq_Hpm1}
a_{\mu}^{\mathrm{NP}}&=a_{\mu}^{H^{\pm}}+a_{\mu}^{W^{\pm}} \nonumber\\
    &=-\frac{4m_{\mu}}{e} \left( \sum_{k=1}^2 \mathrm{Re}[c_{\mu\mu R}^{H^\pm} +c_{\mu\mu R}^{W^{\pm}}]\right).
\end{align}
As neutrino masses and interactions enter the muon $g-2$, numerical computations require a set of parameters consistent with the SM phenomenology. To match the experimental values of the neutrino squared mass differences, SM lepton masses and PMNS matrix elements, we performed a numerical exploration with a Monte Carlo (MC) scanning in the next section.

\section{Results}
 
\begin{figure}[h]
    \centering
    \includegraphics[scale=0.65]{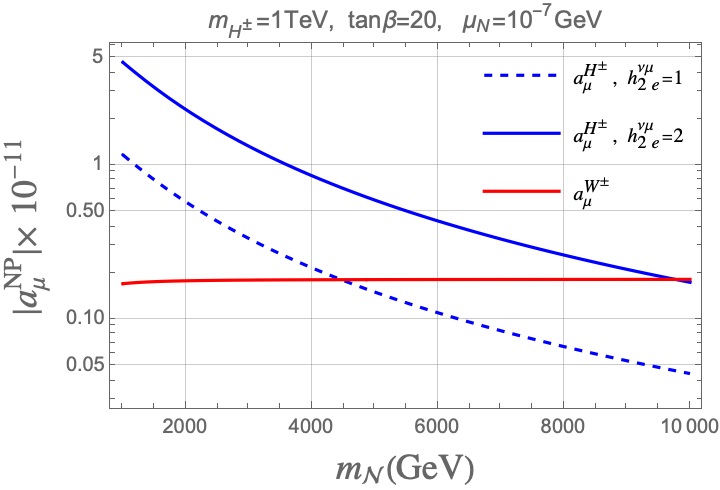}
    \caption{Absolute value contributions to the total muon anomalous magnetic moment $a_\mu$ as a function of the degenerated Majorana neutrino masses $m_{\mathcal{N}}$. Charged Higgs boson (blue) and $W$ boson (red) contributions in the one-loop diagrams of Fig.\ref{g-2feynman} are shown for the displayed fixed parameters. All other Yukawa couplings are set to zero except for $h_{2e}^{\nu \mu}$. }
    \label{g-2_W_H}
\end{figure}

We show in Fig.~\ref{g-2_W_H} the absolute value of the one-loop contributions to $a_{\mu}$ from the $W^{\pm}-\mathcal{N}$ and $H^{\pm}-\mathcal{N}$ loops in Fig.~\ref{g-2feynman} (c) and (d) as a function of the heavy Majorana neutrino $m_{\mathcal{N}}$. We can see that the charged Higgs boson contribution (blue lines) dominates for low TeV Majorana neutrino masses, whereas for very heavy $m_{\mathcal{N}}$ the dominant one comes from the $W$ boson loop when reaching the asymptotic limit of the loop function $\tilde{f}_V(m_{\mathcal{N}}/m_W)$, $-1/6$, when $m_{\mathcal{N}}\to \infty$ (red line). However, the size of the $W$ boson contribution is around two orders of magnitude smaller than the $H^{\pm}$ loop and gets a negative sign when entering the definition of $\Delta a_{\mu}$, meaning that the MC scan will prefer positive contributions to $\Delta a_{\mu}$. This can be explained by looking at the Lagrangian in Eq.(\ref{W_Lagrangian}) where we see that the contribution is inversely proportional to the $\mu_N$ scale which should be of order $\mathrm{keV}$, smaller values could give a larger $W$ boson contribution but would not be consistent with the observed squared mass neutrino mass differences. This means that the largest possible contribution to the total $a_\mu$ at one-loop level will come from the charged Higgs loops for low TeV masses, $a_{\mu}^{H^\pm}\approx5\times10^{-11}$ for $\mathcal{O}(1)$ Yukawa couplings. This also indicates that the MC routine might prefer high and intermediate values for $m_{H^{\pm}}$, regardless of the scale $\mu_N$.

We define the MC routine using the BMW value for $\Delta a_{\mu}$ and the current 90$\%$ C.L. limits for $\mu\to e\gamma$ \cite{MEG:2016leq}, $\tau\to e\gamma$ \cite{BaBar:2009hkt}  and $\tau\to \mu\gamma$ \cite{Belle:2021ysv} from MEG, BaBar and Belle respectively. The parameter space is scanned in the following ranges using the normal ordering scheme for the neutrino squared mass differences\footnote{We found negligible differences when using the inverted ordering scheme.},

\begin{align}m_{\mathcal{N}},\,m_{H^{\pm}}\in[0.5,\,10]\mathrm{TeV} & ,\qquad\tan\beta\in[5,\,50],\nonumber\\
\mu_{N}\in[1\times10^{-8},\,1\times10^{-4}],\quad & \mathrm{{Re,\,Im}(}h_{2e}^{\nu\mu(\tau)})\in[-3,\,3],\nonumber\\
h_{2\mu}^{\nu e},\,h_{2\mu}^{\nu\mu},\,h_{2\mu}^{\nu\tau}\in[-3,\,3] & ,
\end{align}

where the values for the Yukawa couplings are taken within the perturbative bounds of $\sqrt{4\pi}\sim\pm3$.

\begin{figure}[h]
    \centering
    \includegraphics[scale=0.5]{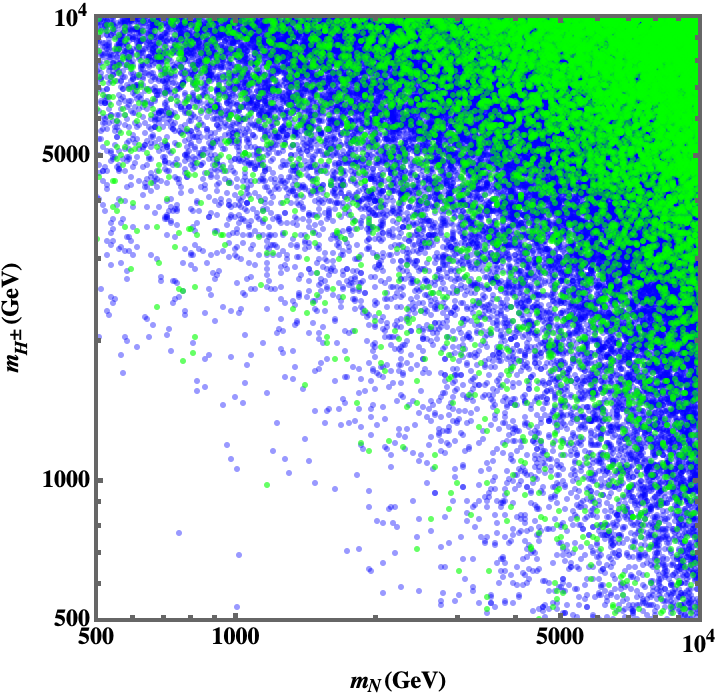}
    \caption{Allowed masses for the charged Higgs boson and exotic neutrinos compatible with muon $g-2$ at 68 (95) \% C.L in green (blue).}
    \label{g-2_mH_vs_mN}
\end{figure}

We first present in Fig.~\ref{g-2_mH_vs_mN} the parameter region where the total $g-2$ contribution matches the experimental value at 95\% C.L. in the $m_{H^{\pm}}$ vs $m_{\mathcal{N}}$ plane. The MC scan shows a preference for both masses at $10$ TeV and sets a lower bound of around $1.5$ TeV, which is consistent with $b\to s\gamma$ constraints on $m_{H^{\pm}}$ for large $\tan\beta\approx m_t/m_b$ \cite{Misiak:2017bgg}.

\begin{figure}[h]
    \centering
    \includegraphics[scale=0.5]{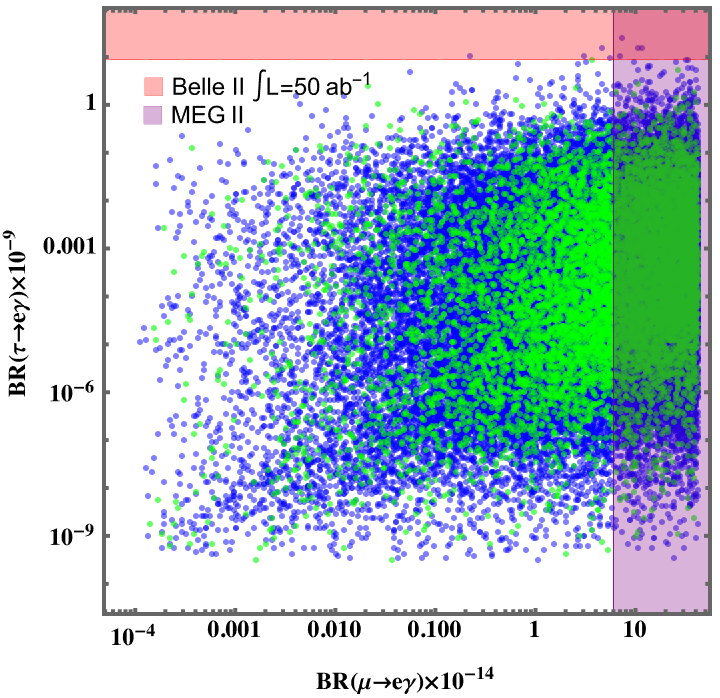}
    \caption{Projected 1 (green) and 2$\sigma$ (blue) regions for the branching ratios of the charged lepton flavor violating decays $\tau\to e\gamma$ and $\mu\to e\gamma$. The red and light purple regions are the 
the Belle II \cite{Belle-II:2018jsg} and MEG II \cite{MEGII:2018kmf} future sensitivities respectively}
\label{fig:tauegamma_muegamma}
\end{figure}

In Fig.\ref{fig:tauegamma_muegamma} we show the predictions for the branching ratios of the cLFV decays $\tau\to e\gamma$ and $\mu\to e \gamma$. The MC generates points for the model that simultaneously respect the current bounds at 90$\%$ C.L., with a large portion of the $1\sigma$ samples within reach of the future sensitivity at MEG II. Regarding the Belle II projections for $\tau\to e\gamma$, the model prediction is two orders of magnitude below.

\begin{figure}[h]
    \centering
    \includegraphics[scale=0.5]{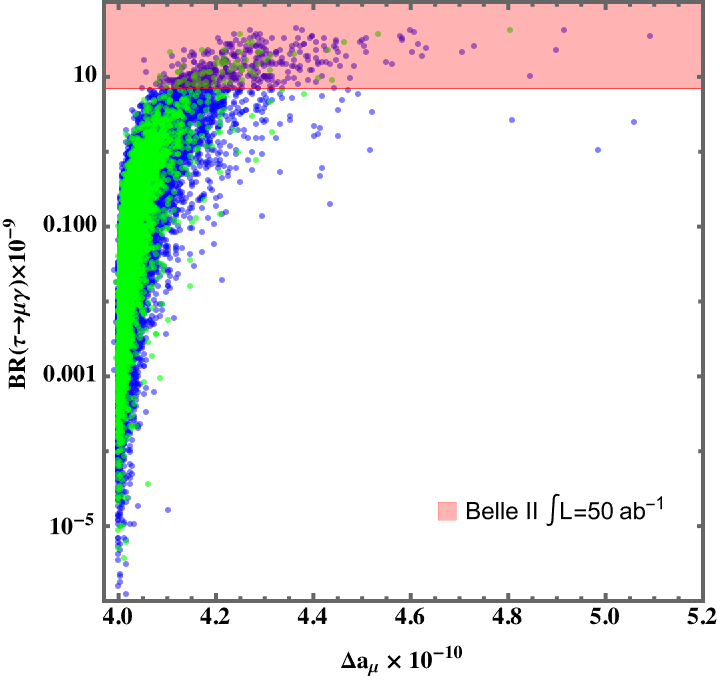}
    \caption{Projected 1 (green) and 2$\sigma$ (blue) regions for the branching ratio of the $\tau\to \mu\gamma$ decays vs the total $\Delta a_{\mu}$. The red region is the
the Belle II \cite{Belle-II:2018jsg} future sensitivity.}
\label{fig:taumugamma_deltaa}
\end{figure}

However, the foreseen sensitivity of the Belle II experiment could prove the $\tau\to \mu\gamma$ phase space within the $1\sigma$ level (Fig. \ref{fig:taumugamma_deltaa}), although the majority of points in the $1\sigma$ region lie well below the Belle II projection. Regarding the total $\Delta a_\mu$, the model predicts SM-like values as expected, all within the $1\sigma$ region with a maximum possible value of $\Delta a_{\mu}\sim5\times10^{-10}$. We note that the majority of samples are generated towards the upper limit of the experimental uncertainty, consistent with the dominance of the charged Higgs contributions. We can also see a few points below the central value of $\Delta a_\mu=4.0\times10^{-10}$ at the 2 $\sigma$ level, resulting from the (subdominant) contributions of the $W$ boson. 

\begin{figure}[h]
    \centering
    \includegraphics[scale=0.5]{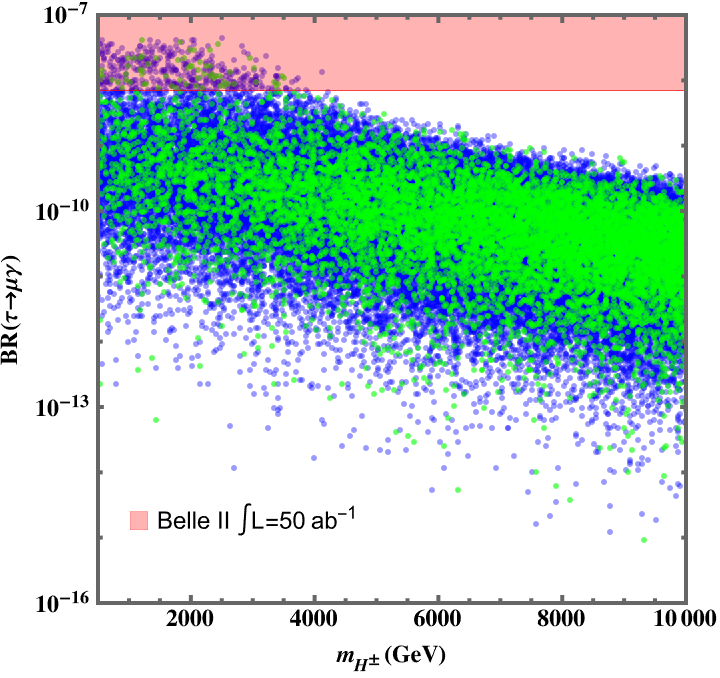}
    \caption{Projected 1 (green) and 2$\sigma$ (blue) regions for the branching ratio of the cLFV decay $\tau\to \mu\gamma$ as a function of $m_{H^{\pm}}$. The red region is
the Belle II \cite{Belle-II:2018jsg} future sensitivity.}
\label{fig:tamugamma_mH}
\end{figure}

Finally, Fig. \ref{fig:tamugamma_mH} indicates that Belle II could prove (or exclude) the $1$-$4$ TeV range for the charged Higgs boson mass. With respect to the other model parameters, we find no preference for either the $\tan\beta$ or Yukawa couplings, except for the $h_{2\mu}^{\nu e}$ and $h_{2\mu}^{\nu \mu}$ couplings, which the MC sampling shows are compatible with null values at 95$\%$ C.L.. The independence of $\tan\beta$ can be understood from the $\lambda_{ai}^{L,R}$ products in the $c_{(ab)R}^{H^{\pm}}$ couplings. The $\tan^2\beta$ factor in $\lambda_{ai}^{L}\lambda_{ai}^{L}$ is suppressed by the $U_{\mathrm{PMNS}}(\hat{m}_{\nu}/\mu_N)U_{\mathrm{PMNS}}^\dagger$ factor, $\tan\beta$ cancels in the $\lambda_{ai}^{L}\lambda_{ai}^{R}$ term and finally, it gets suppressed as $\tan^{-2}\beta$ in the $\lambda_{ai}^{R}\lambda_{ai}^{R}$ factor.

\section{Conclusions}\label{conclusions}

We presented a gauged non-universal $U(1)_X$ extension of the 2HDM capable of explaining the mass hierarchy of charged fermions and neutrino oscillations via the non-universality of the new quantum numbers and a $\mathbb{Z}_{2}$ symmetry. In the context of the most recent BMW lattice calculations of the muon $g-2$, we analyzed the one-loop contributions focusing on the effects of heavy Majorana neutrinos and charged Higgs bosons. Finding that the dominant contribution to \(a_\mu\) arises from the \(H^\pm\)-\(\mathcal{N}\) loop for low TeV-scale Majorana neutrino masses, reaching \(a_\mu^{H^\pm} \approx 5 \times 10^{-11}\) for \(\mathcal{O}(1)\) Yukawa couplings. In contrast, the contribution of the \(W\)-\(\mathcal{N}\) loop is suppressed by two orders of magnitude and carries a negative sign when entering $\Delta a_{\mu}$. We performed a MC scan of the parameter space, constrained by experimental limits on \(\mu \to e\gamma\), \(\tau \to e\gamma\), and \(\tau \to \mu\gamma\), revealing a preference for charged Higgs and Majorana neutrino masses near 10 TeV, with a lower bound of approximately 1.5 TeV. The model predicts branching ratios for the cLFV decays that are mostly below current experimental sensitivities. However, a subset of parameter space falls within the projected reach of MEG II for \(\mu \to e\gamma\), while \(\tau \to \mu \gamma\) could be probed by Belle II for intermediate masses of $H^\pm$ in the range (\(1\)-\(4\) TeV). Finally, the total \(\Delta a_\mu\) aligns with the observed values, predominantly driven by \(H^\pm\) contributions. No strong correlations emerge for \(\tan \beta\) or most Yukawa couplings, except for \(h_{2\mu}^{\nu e}\) and \(h_{2\mu}^{\nu \mu}\), which are consistent with zero at 95\% C.L.

\begin{acknowledgements}

R.M. is thankful to Nanjing Normal University for its hospitality during the beginning of this project. The work of C. S. is supported in part by the Excellent Postdoctoral Program of Jiangsu Province Grant No. 2023ZB891. The work of J. S. is supported by the white program of the Physics Graduate School (Programme blanc GS Physique) of the PHENIICS doctoral school [Contract No. D22-ET11].

\end{acknowledgements}

\bibliographystyle{spphys}    
\bibliography{References}   

\end{document}